\documentclass[twocolumn,showpacs,preprintnumbers,amsmath,amssymb]{revtex4}

\usepackage{graphicx}
\usepackage{dcolumn}
\usepackage{bm}

\begin{document}
\preprint{APS/123-QED}
\title{Doubly Special Relativity with a minimum speed and the Uncertainty Principle} 
\author{Cl\'audio Nassif\\
        (e-mail: cnassif@cbpf.br)}
 \altaffiliation{CBPF: Centro Brasileiro de Pesquisas F\'isicas, Rua Dr. Xavuer Sigaud 150, Urca, 22.290-180, Rio de Janeiro-RJ, Brazil.}

\date{\today}

\begin{abstract}
 The present work aims to search for an implementation of a new symmetry in the space-time by introducing the idea of an invariant minimum
 speed scale ($V$). Such a lowest limit $V$, being unattainable by the particles, represents a fundamental and preferred reference frame
 connected to a universal background field (a vacuum energy) that breaks Lorentz symmetry. So there emerges a new principle of symmetry in
 the space-time at the subatomic level for very low energies close to the background frame ($v\approx V$), providing a fundamental
 understanding for the uncertainty principle, i.e., the uncertainty relations should emerge from the space-time with an invariant
 minimum speed.
\end{abstract}



\maketitle
\section{\label{sec:level1} Introduction}

 Einstein\cite{1} criticized the existence of the luminiferous
 ether defended by Lorentz\cite{2}, Fitzgerald\cite{3} and Poincar\'e\cite{4}. So
 he solved the incompatibility between the laws of motion in the
 newtonian mechanics and the laws of electromagnetic fields.

   During the last 30 years of his life, Einstein attempted to bring the
 principles of Quantum Mechanics (QM) and Electromagnetism (EM)
 into General Relativity (GR) by means of a unified field theory\cite{5}\cite{6}\cite{7}. 
 Unfortunately his unification program was not successful in establishing a consistent theory between QM, EM and GR.

 Inspired by the seductive search for new fundamental symmetries in Nature, we attempt to implement a
 uniform background field into the flat space-time. Such a background field connected to a uniform vacuum energy density represents a
 preferred reference frame (an ultra-referential),
 which leads us to postulate a universal and invariant minimum limit of speed for particles with very large wavelengths (very low
 energies). Such a minimum speed that breaks Lorentz symmetry\cite{8} leads us to build a new relativistic
 dynamics\cite{9}\cite{10} that provides a fundamental understanding of the quantum uncertainties.

 An interesting motivation for considering the idea of a minimum speed $V$ in the quantum space-time is that, even if only in the field 
 of likely hypotheses for the purpose of understanding the
 whole theory, there is a privileged background frame connected to a vacuum energy. This should be accepted as a postulate, from which
 emerge with clarity and authenticity new concepts, ideas and physical greatnesses. The very idea of an unattainable minimum speed is the
 first consequence of the existence of a background frame, and it relates to the unattainable temperature of absolute zero of the third law
 of thermodynamics that prevents it from being reached(see ref.\cite{10}). Thus we must admit that the idea of a background field
 should not be neglected, as it exists potentially. And so we must emphasize that the ideal condition for Bose-Einstein condensation
 ($T=0K$) is the arrival point on the path toward the background field where the particles lose completely their identity, i.e.,
 we have a complete
 delocalization when $v\rightarrow V$ close to the absolute zero temperature\cite{10}. Moreover, such an idea of a background
 field (an ultra-referential) connected to an
 invariant minimum speed was successfully applied in determining the tiny value of the cosmological constant and the
 vacuum energy density, being in good agreement with recent observations\cite{9}. And here we mention that
 the idea of a background field represented by the vacuum or cosmological constant has been used to explain the generation of mass for all
 bodies\cite{11}\cite{12}.

   As the resting state is forbidden for the photon ($v=c$), which behaves like a wave-particle, and knowing that a massive particle 
 is also a wave-particle according to the de-Broglie reasoning of reciprocity, so by expanding such a reasoning to obtain a stronger 
 symmetry, we are led to believe that the resting state is also forbidden for a massive particle ($v<c$). Thus the space-time with a
 non-zero minimum speed $V$ and the maximum speed $c$ reveals a fundamental symmetry, which is necessary to ensure the wave-particle 
 duality for both light and matter. For this reason, the present doubly special relativity was denominated as Symmetrical Special 
 Relativity (SSR)\cite{9}\cite{10}.

 The dynamics of particles in the presence of a universal (privileged) background reference frame connected to $V$ is
 within the context of ideas of Mach\cite{13}, Schr\"{o}dinger\cite{14} and Sciama\cite{15}, where there should be an absolute inertial
 reference frame in relation to which we have the inertia of all moving bodies. However, we must emphasize that the approach we intend to
 use is not classical as the Machian ideas since the lowest limit of speed $V$ plays the role of a preferred reference frame
 of background field instead of the inertial frame of fixed stars.

 It is very interesting to notice that the idea of a universal background field was sought in vain by
 Einstein\cite{16}\cite{17}\cite{18}\cite{19}\cite{20}\cite{21}, motivated firstly by Lorentz. It was Einstein who coined the
 term {\it ultra-referential} as the fundamental aspect of reality for representing a universal background field\cite{22}. Based on such a concept, let us call {\it ultra-referential} $S_V$ to be
 the universal background field of a fundamental (preferred) reference frame connected to $V$.

 Since there are alternative relativity theories that are much discussed in the literature, we quote the DSR theories that were first
proposed by Camelia\cite{23}\cite{24}\cite{25}\cite{26}. Other alternative theories for DSR were also proposed later by Magueijo, Smolin
and Albrecht\cite{27}\cite{28}\cite{29}. Still another extension of Special Relativity (SR) as triply special relativity was
proposed\cite{30}. And the quantizing speeds with the cosmological constant should be also mentioned\cite{31}.

\section{\label{sec:level1} Reference frames and space-time interval in SSR}

   Since we cannot think about a reference system constituted by a set of infinite points at rest in the quantum space-time of
 SSR\cite{9}\cite{10}, we should define a non-Galilean reference system essentially as a set of all the particles having the same state of
 movement (speed $v$) with respect to the ultra-referential $S_V$ (background frame) so that $v>V$. Hence, SSR should contain three
postulates, namely:

 1)-the constancy of the speed of light ($c$);

 2)-the non-equivalence (asymmetry) of the non-Galilean reference frames due to the background frame $S_V$ that breaks Lorentz symmetry,
  i.e., we cannot exchange $v$ for $-v$ by the inverse transformations, since $``v-v">V$ (see ref.\cite{9});

 3)-the covariance of the ultra-referential $S_V$ (background frame) connected to an invariant and unattainable minimum speed $V$ in the
subatomic world.

Let us assume the reference frame $S^{\prime}$ with a speed $v$ in relation to the ultra-referential $S_V$ according to figure 1.

\begin{figure}
\includegraphics[scale=0.8]{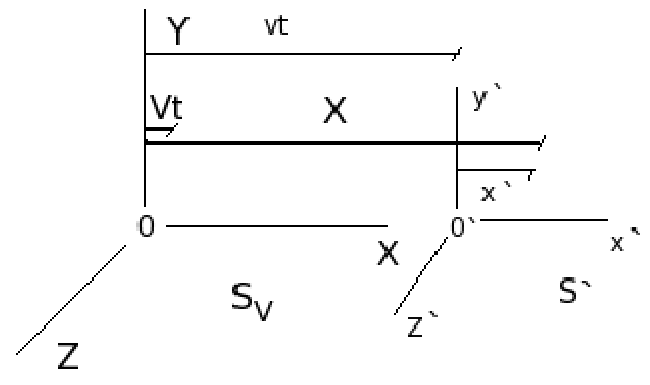}
\caption{$S^{\prime}$ moves with a speed $v$ with respect to the background field of the covariant ultra-referential $S_V$.
 If $V=0$, $S_V$ is eliminated (no vacuum energy) and thus the Galilean frame $S$ takes place, recovering Lorentz
 transformations.}
\end{figure}

Hence, consider the motion at one spatial dimension, namely $(1+1)D$ space-time with background field-$S_V$. So we write the following
 transformations:

  \begin{equation}
 x^{\prime}=\Psi(X-\beta_{*}ct)=\Psi(X-vt+Vt),
  \end{equation}
where $\beta_{*}=\beta\epsilon=\beta(1-\alpha)$, being $\beta=v/c$ and $\alpha=V/v$, so that $\beta_{*}\rightarrow 0$ for $v\rightarrow V$
 or $\alpha\rightarrow 1$.

 \begin{equation}
 t^{\prime}=\Psi(t-\frac{\beta_{*}X}{c})=\Psi(t-\frac{vX}{c^2}+\frac{VX}{c^2}),
  \end{equation}
being $\vec v=v_x{\bf x}$. We have $\Psi=\frac{\sqrt{1-\alpha^2}}{\sqrt{1-\beta^2}}$. If we make $V\rightarrow 0$ ($\alpha\rightarrow 0$),
 we recover Lorentz transformations.

  In order to get the transformations (1) and (2), let us consider the following more general transformations:
$x^{\prime}=\theta\gamma(X-\epsilon_1vt)$ and
 $t^{\prime}=\theta\gamma(t-\frac{\epsilon_2vX}{c^2})$, where $\theta$, $\epsilon_1$ and $\epsilon_2$ are factors (functions) to be
 determined. We hope all these
 factors depend on $\alpha$, such that, for $\alpha\rightarrow 0$ ($V\rightarrow
 0$), we recover Lorentz transformations as a particular case
 ($\theta=1$, $\epsilon_1=1$ and $\epsilon_2=1$). By using those transformations to perform
$[c^2t^{\prime 2}-x^{\prime 2}]$, we find the identity: $[c^2t^{\prime
 2}-x^{\prime 2}]=
\theta^2\gamma^2[c^2t^2-2\epsilon_1vtX+2\epsilon_2vtX-\epsilon_1^2v^2t^2+\frac{\epsilon_2^2v^2X^2}{c^2}-X^2]$.
 Since the metric tensor is diagonal, the crossed terms must vanish and so we assure that
$\epsilon_1=\epsilon_2=\epsilon$. Due to this fact, the crossed terms
 ($2\epsilon vtX$) are cancelled between themselves and finally we obtain $[c^2t^{\prime 2}-x^{\prime 2}]=
 \theta^2\gamma^2(1-\frac{\epsilon^2 v^2}{c^2})[c^2t^2-X^2]$. For
 $\alpha\rightarrow 0$ ($\epsilon=1$ and
$\theta=1$), we reinstate $[c^2t^{\prime 2}-x^{\prime 2}]=[c^2t^2-x^2]$ of SR. Now we write the following
transformations: $x^{\prime}=\theta\gamma(X-\epsilon
 vt)\equiv\theta\gamma(X-vt+\delta)$ and
$t^{\prime}=\theta\gamma(t-\frac{\epsilon
 vX}{c^2})\equiv\theta\gamma(t-\frac{vX}{c^2}+\Delta)$, where we assume $\delta=\delta(V)$ and $\Delta=\Delta(V)$, so that $\delta
 =\Delta=0$ for $V\rightarrow 0$, which implies $\epsilon=1$.
 So from such transformations we extract: $-vt+\delta(V)\equiv-\epsilon vt$ and
$-\frac{vX}{c^2}+\Delta(V)\equiv-\frac{\epsilon vX}{c^2}$, from where we obtain
 $\epsilon=(1-\frac{\delta(V)}{vt})=(1-\frac{c^2\Delta(V)}{vX})$. As
 $\epsilon$ is a dimensionless factor, we immediately conclude that $\delta(V)=Vt$ and
 $\Delta(V)=\frac{VX}{c^2}$, so that we find
$\epsilon=(1-\frac{V}{v})=(1-\alpha)$. On the other hand, we can determine $\theta$ as follows: $\theta$ is a function of $\alpha$ ($\theta(\alpha)$), such
 that
 $\theta=1$ for
 $\alpha=0$, which also leads to $\epsilon=1$ in order to recover Lorentz transformations. So, as $\epsilon$ depends on
 $\alpha$, we conclude that $\theta$ can also be expressed in terms of $\epsilon$, namely
 $\theta=\theta(\epsilon)=\theta[(1-\alpha)]$, where
$\epsilon=(1-\alpha)$. Therefore we can write
 $\theta=\theta[(1-\alpha)]=[f(\alpha)(1-\alpha)]^k$, where the exponent $k>0$.

 The function $f(\alpha)$ and $k$ will be estimated by satisfying the following conditions:

i) as $\theta=1$ for $\alpha=0$ ($V=0$), this implies $f(0)=1$.

ii) the function $\theta\gamma =
\frac{[f(\alpha)(1-\alpha)]^k}{(1-\beta^2)^{\frac{1}{2}}}=\frac{[f(\alpha)(1-\alpha)]^k}
{[(1+\beta)(1-\beta)]^{\frac{1}{2}}}$ should have a symmetrical behavior, that is to say it approaches to zero when closer to $V$ ($\alpha\rightarrow 1$), and
 in the same way to the infinite when closer to $c$ ($\beta\rightarrow 1$). In other words, this means that the numerator of the function $\theta\gamma$, which
 depends on $\alpha$ should have the same shape of its denominator, which depends on $\beta$. Due to such conditions, we naturally conclude that $k=1/2$ and
$f(\alpha)=(1+\alpha)$, so that $\theta\gamma =
\frac{[(1+\alpha)(1-\alpha)]^{\frac{1}{2}}}{[(1+\beta)(1-\beta)]^{\frac{1}{2}}}=
\frac{(1-\alpha^2)^{\frac{1}{2}}}{(1-\beta^2)^\frac{1}{2}}=\frac{\sqrt{1-V^2/v^2}}{\sqrt{1-v^2/c^2}}=\Psi$,
 where $\theta =(1-\alpha^2)^{1/2}=(1-V^2/v^2)^{1/2}$.

The transformations (1) and (2) are the direct transformations from $S_V$ [$X^{\mu}=(X,ict)$] to $S^{\prime}$
 [$x^{\prime\nu}=(x^{\prime},ict^{\prime})$], where we have $x^{\prime\nu}=\Omega^{\nu}_{\mu} X^{\mu}$ ($x^{\prime}=\Omega X$), so that we
 obtain the following matrix of transformation:

\begin{equation}
\displaystyle\Omega=
\begin{pmatrix}
\Psi & i\beta (1-\alpha)\Psi \\
-i\beta (1-\alpha)\Psi & \Psi
\end{pmatrix},
\end{equation}
such that $\Omega\rightarrow\ L$ (Lorentz matrix of rotation) for
 $\alpha\rightarrow 0$ ($\Psi\rightarrow\gamma$). We should investigate whether the transformations (3) form a group. However, such an
 investigation can form the basis of a further work.

We obtain $det\Omega
 =\frac{(1-\alpha^2)}{(1-\beta^2)}[1-\beta^2(1-\alpha)^2]$, where $0<det\Omega<1$. Since
$V$ ($S_V$) is unattainable ($v>V)$, this assures that $\alpha=V/v<1$
 and therefore the matrix $\Omega$
admits inverse ($det\Omega\neq 0$ $(>0)$). However $\Omega$ is a non-orthogonal matrix
($det\Omega\neq\pm 1$) and so it does not represent a rotation matrix
 ($det\Omega\neq 1$) in such a space-time due to the presence of the privileged frame of background field $S_V$ that breaks strongly the
 invariance of the
 norm of the 4-vector of SR. Actually such an effect ($det\Omega\approx 0$ for $\alpha\approx 1$) emerges from a new relativistic physics
 of SSR
 for treating much lower energies at ultra-infrared regime closer to $S_V$ (very large wavelengths).

 We notice that $det\Omega$ is a function of the speed $v$ with respect to $S_V$. In the approximation for $v>>V$ ($\alpha\approx 0$), we obtain
 $det\Omega\approx
 1$ and so we practically reinstate the rotational behavior of Lorentz matrix as a particular regime for higher energies.
 If we make $V=0$ ($\alpha=0$), we recover $det\Omega=1$.

The inverse transformations (from $S^{\prime}$ to $S_V$) are

 \begin{equation}
 X=\Psi^{\prime}(x^{\prime}+\beta_{*}ct^{\prime})=\Psi^{\prime}(x^{\prime}+vt^{\prime}-Vt^{\prime}),
  \end{equation}

 \begin{equation}
 t=\Psi^{\prime}\left(t^{\prime}+\frac{\beta_{*}
 x^{\prime}}{c}\right)=\Psi^{\prime}\left(t^{\prime}+\frac{vx^{\prime}}{c^2}-\frac{Vx^{\prime}}{c^2}\right).
  \end{equation}

In matrix form, we have the inverse transformation
 $X^{\mu}=\Omega^{\mu}_{\nu} x^{\prime\nu}$
 ($X=\Omega^{-1}x^{\prime}$), so that the inverse matrix is

\begin{equation}
\displaystyle\Omega^{-1}=
\begin{pmatrix}
\Psi^{\prime} & -i\beta (1-\alpha)\Psi^{\prime} \\
 i\beta (1-\alpha)\Psi^{\prime} & \Psi^{\prime}
\end{pmatrix},
\end{equation}
where we can show that $\Psi^{\prime}$=$\Psi^{-1}/[1-\beta^2(1-\alpha)^2]$, so that we must satisfy $\Omega^{-1}\Omega=I$.

 Indeed we have $\Psi^{\prime}\neq\Psi$ and therefore
 $\Omega^{-1}\neq\Omega^T$. This non-orthogonal aspect of
$\Omega$ has an important physical implication. In order to understand such an implication, let us first consider the orthogonal (e.g: rotation) aspect of
 Lorentz matrix in SR. Under SR, we have $\alpha=0$, so that
 $\Psi^{\prime}\rightarrow\gamma^{\prime}=\gamma=(1-\beta^2)^{-1/2}$.
  This symmetry ($\gamma^{\prime}=\gamma$, $L^{-1}=L^T$) happens because the galilean reference frames allow us to exchange the speed $v$ (of $S^{\prime}$)
 for $-v$ (of $S$) when we are at rest at
$S^{\prime}$. However, under SSR, since there is no rest at
 $S^{\prime}$, we cannot exchange $v$ (of $S^{\prime}$) for $-v$ (of $S_V$)
due to that asymmetry ($\Psi^{\prime}\neq\Psi$,
 $\Omega^{-1}\neq\Omega^T$). Due to this fact,
$S_V$ must be covariant, namely $V$ remains invariant for any change of reference frame in such a space-time. Thus we can notice that the 
paradox of twins,
 which appears due to the symmetry by exchange of $v$ for $-v$ in SR should be naturally eliminated in SSR\cite{9}\cite{10}, where only
 the reference frame $S^{\prime}$ can move
 with respect to $S_V$. So $S_V$ remains covariant (invariant for any change of reference frame).
  We have $det\Omega=\Psi^2[1-\beta^2(1-\alpha)^2]\Rightarrow
 [(det\Omega)\Psi^{-2}]=[1-\beta^2(1-\alpha)^2]$. So
we can alternatively write
 $\Psi^{\prime}$=$\Psi^{-1}/[1-\beta^2(1-\alpha)^2]=\Psi^{-1}/[(det\Omega)\Psi^{-2}]
=\Psi/det\Omega$. By inserting this result in (6) to replace
 $\Psi^{\prime}$, we obtain the relationship between the inverse matrix and the transposed matrix of $\Omega$, namely $\Omega^{-1}=\Omega^T/det\Omega$. Indeed
 $\Omega$ is a non-orthogonal matrix, since we have $det\Omega\neq\pm 1$.

   According to SSR, the starting point for obtaining the true motion of all the particles at $S^{\prime}$ is the ultra-referential
 $S_V$ (see figure 2). However, due to the non-locality of $S_V$, being unattainable by any particle, the existence of a classical
 observer at $S_V$ becomes inconceivable. Hence, let us think about a non-Galilean frame $S_0$ with a certain intermediary speed
 ($V<<v_0<<c$) in order to represent the new starting point for observing the motion of $S^{\prime}$ (figure 2). At this
 non-Galilean reference frame $S_0$ (for $v=v_0$), which plays the similar role of a ``rest'', we must restore all the newtonian parameters
 of the particles such as the proper time interval $\Delta\tau$, i.e., $\Delta\tau=\Delta t$ for $v=v_0$; the mass $m_0$, i.e.,
 $m(v=v_0)= m_0$\cite{9}\cite{10}, among others. Therefore, in this sense, the frame $S_0$ under SSR plays a role that is similar to the
 frame $S$
 under SR where $\Delta\tau=\Delta t$ for $v=0$, $m(v=0)=m_0$, etc. However we must stress that the classical relative rest ($v=0$) should
 be replaced by a universal ``quantum rest'' $v_0(\neq 0)$ of the non-Galilean reference frame $S_0$.

\begin{figure}
\includegraphics[scale=0.55]{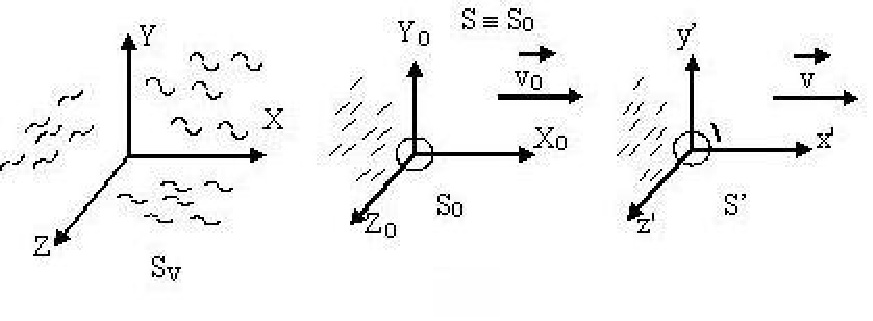}
\caption{As $S_0$ is fixed (universal), being $v_0 (>>V)$ given with respect to $S_V$, we should also consider the interval $V~(S_V)<v
\leq v_0$ ($S_0$). Such interval introduces a new symmetry for the space-time of SSR. Thus we expect that new and interesting results 
take place. For this interval, we have $\Psi(v)\leq 1$.}
\end{figure}

  According to figures 2 and 3, we notice that both $S_V$ and $S_0$ are universal non-Galilean frames, whereas $S^{\prime}$ is a 
rolling non-Galilean frame of the particle moving within the interval of speeds $V<v<c$.

  Since the rolling frame $S^{\prime}$ is a non-Galilean frame due to the impossibility to find a set of points at rest on it, we cannot
 place a particle exactly on its origin $O^{\prime}$ as there should be a delocalization $\Delta
 x^{\prime}$ ($=\overline{O^{\prime}C}$) around the origin $O^{\prime}$ of the frame $S^{\prime}$ (see figure 3). Actually we want to show
 that such
 delocalization $\Delta x^{\prime}$ is a function which should depend on the speed $v$ of $S^{\prime}$ with respect to $S_V$, namely, for
 example, if $S^{\prime}\rightarrow S_V$ ($v\rightarrow V$), we would have $\Delta x^{\prime}\rightarrow\infty$ (complete delocalization)
 due to the non-local aspect of the ultra-referential $S_V$. On the other hand, if $S^{\prime}\rightarrow S_c(v\rightarrow c)$, we would
 have $\Delta x^{\prime}\rightarrow 0$ (much better located on $O^{\prime}$). So let us search for the function $\Delta x^{\prime}=\Delta
 x^{\prime}_v=\Delta x^{\prime}(v)$, starting from figure 3.

\begin{figure}
\includegraphics[scale=0.6]{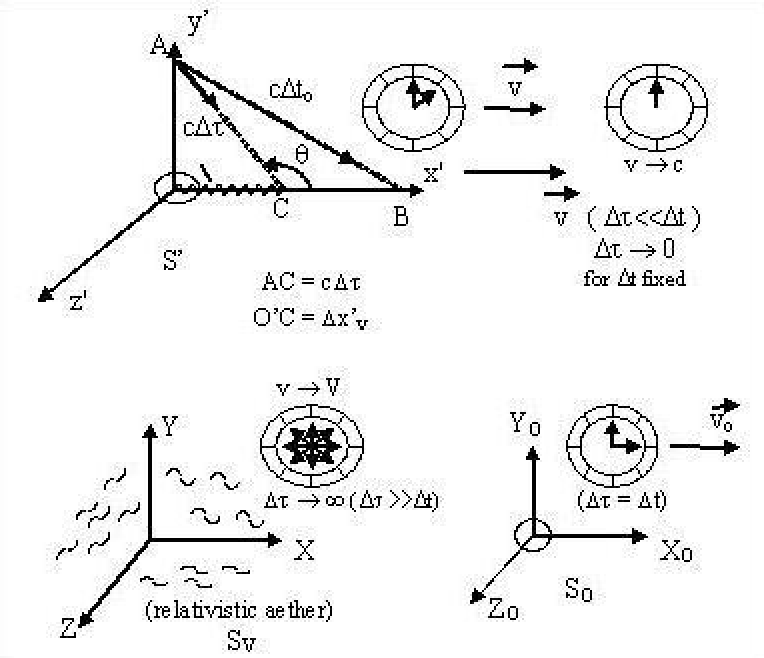}
\caption{We have four imaginary clocks associated to the non-Galilean reference frames $S_0$, $S^{\prime}$, the ultra-referential $S_V$ for
 $V$ and $S_c$ for $c$. We observe that the time (interval $\Delta\tau$) elapses much faster closer to infinite
 ($\Delta\tau\rightarrow\infty$) when one approximates to $S_V$ and, on the other hand, it tends to stop ($\Delta\tau\rightarrow 0$) when
 $v\rightarrow c $, providing a strong symmetry for SSR.}
\end{figure}

  At the reference frame $S^{\prime}$ in figure 3, let us consider a photon emitted from a point $A$ at $y^{\prime}$, in the direction
 $\overline {AO^{\prime}}$, which occurs only if $S^{\prime}$ were a Galilean reference frame. However, since the electron cannot be
 thought of as a point at rest on its proper non-Galilean reference frame $S^{\prime}$ and cannot be located exactly on $O^{\prime}$, then
 its delocalization $\overline {O^{\prime}C}$ ($=\Delta x^{\prime}_v$) causes the photon to deviate from the direction $\overline
 {AO^{\prime}}$ to $\overline {AC}$.  Hence, instead of simply the segment $\overline {AO^{\prime}}$, a rectangle triangle $AO^{\prime}C$
 is formed at the proper non-Galilean frame $S^{\prime}$ where it is not possible to find a set of points at rest.

   From the non-Galilean frame $S_0$ ($v=v_0$), which plays a similar role of the improper Galilean frame $S$ ($v=0$), we see the path
$\overline {AB}$ of the photon (figure 3). Hence the rectangle triangle $AO^{\prime}B$ is formed. Since the vertical leg $\overline {AO^{\prime}}$
 is common to the triangles $AO^{\prime}C$ ($S^{\prime}$) and $AO^{\prime}B$ ($S_0$), we have

 \begin{equation}
(\overline {AO^{\prime}})^2=(\overline {AC})^2-(\overline {O^{\prime}C})^2=
(\overline {AB})^2 - (\overline {O^{\prime}B})^2,
\end{equation}
 that is

 \begin{equation}
(c\Delta\tau)^2-(\Delta x^{\prime}_v)^2=(c\Delta t)^2 - (v\Delta t)^2,
\end{equation}
where $\Delta t=\Delta t_0$ (improper time at $S_0$), being $S_0$ the improper non-Galilean frame in SSR. So, from (8) we expect that, 
for the case $v=v_0$ (at $S_0$), we have $\Delta\tau=\Delta t$, leading to $\Delta x^{\prime}_{[v=v_0]}=v_0\Delta\tau$ (see (8)).

  As $\Delta x^{\prime}_v$ is a function of $v$, being an ``internal displacement" (delocalization) given on the proper non-Galilean
 frame $S^{\prime}$, we may write it in the following way: 

\begin{equation}
\Delta x^{\prime}_v= f(v)\Delta\tau,
\end{equation}
 where $f(v)$ is a function of $v$. It has also dimension of speed, however it could be thought as if it were a kind of
{\it ``internal motion" $v_{int}$} given in a reciprocal space of momentum for representing the delocalization $\Delta x^{\prime}_v$
on position of the particle at its proper non-Galilean reference frame $S^{\prime}$. So we have $\Delta x^{\prime}_v=v_{int}\Delta\tau$,
where $f(v)(=v_{int})$ is the reciprocal speed ($v_{rec}$), which represents the rate of change of delocalization in the 
time $\tau$. Here we draw attention to the fact that such delocalization $\Delta x^{\prime}_v$ is not merely a spatial displacement 
of a particle, since it represents a kind of stretching of the ``particle" itself which works like a non-punctual object. This
stretching has no classical analogy and it leads to an internal motion, which is given by the reciprocal speed. So, when $v$ tends to $V$, the ``particle"
is no longer as a punctual object in order to become completely non-localized due to the spatial dilation. We will compute such an effect
connected to $v_{rec}$ and later we will point to the fact that the delocalization $\Delta x^{\prime}_v$ appears as an uncertainty 
on position when viewed from any Galilean reference frame.

 In figure 3 we can see the delocalization $\Delta x^{\prime}_v$ given by the segment $\overline{O^{\prime}C}$ at
the frame $S^{\prime}$, i.e., we have $\Delta x^{\prime}_v=\overline{O^{\prime}C}$. This leads us to think that there is an
uncertainty $\Delta x^{\prime}_v$ on position of the particle, as we will see later. So inserting (9) into (8), we obtain

\begin{equation}
\Delta\tau\left[1-\frac{[f(v)]^2}{c^2}\right]^{\frac{1}{2}}=\Delta t\left(1-\frac{v^2}{c^2}\right)^{\frac{1}{2}},
\end{equation}
 where $f(v)=v_{int}=v_{rec}=v_{reciprocal}$.

  As we have $v\leq c$, we should find $f(v)\leq c$ in order to prevent an imaginary number in the 1st.member of (10).

 The domain of $f(v)$ is such that $V\leq v\leq c$. Thus let us also admit that its image is $V\leq f(v)\leq c$ since $f(v)$ has dimension
 of speed for representing $v_{rec}$, which should also be limited by $V$ and $c$.

  Let us make $[f(v)]^2/c^2 = f^2/c^2 =v_{int}^2/c^2=\alpha^2$, whereas we already know that $v^2/c^2 =\beta^2$. Thus, from (10) we
have the following cases:

- (i) When $v\rightarrow c$ ($\beta\rightarrow\beta_{max}=1$), the relativistic correction in its 2nd member (right-hand side) prevails,
 while the correction on the left-hand side becomes practically neglectable, i.e., we should have $v_{int}=f(v)<<c$, where
 $lim._{v\rightarrow c}f(v)=f_{min}=(v_{int})_{min}=V$ ($\alpha\rightarrow\alpha_{min}=V/c=\xi<<1$). So, in this case, we find 
 $\Delta t>>\Delta\tau$.

 -(ii) On the other hand, due to the idea of symmetry, if $v\rightarrow V$ ($\beta\rightarrow\beta_{min}=V /c =\xi$), there is no
 substantial relativistic correction on the right-hand side of (10), while the correction on the left-hand side becomes now considerable,
 that is, we should have $lim._{v\rightarrow V}f(v)=f_{max}=(v_{int})_{max}=c$ ($\alpha\rightarrow\alpha_{max}=1$). So, in this case,
 we find $\Delta\tau>>\Delta t$.

 In short, from (i) and (ii) we observe that, if $v\rightarrow v_{max}=c$, then $f\rightarrow f_{min}=(v_{int})_{min}=V$ and, if
 $v\rightarrow v_{min}=V$, then $f\rightarrow f_{max}=(v_{int})_{max}=c$. So now we indeed perceive that the ``internal motion" $v_{int}$
 works like a reciprocal speed $v_{rec}$ leading to the delocalization $\Delta x^{\prime}_v$. In other words, we say that, when
 the speed $v$ increases to $c$, the reciprocal one ($v_{rec}$) decreases to $V$. On the other hand, when $v$ tends to $V$ ($S_V$), so
 $v_{rec}$  tends to $c$ leading to a very large delocalization $\Delta x^{\prime}_v$. Due to this fact, we reason that

\begin{equation}
f(v)=v_{int}=v_{rec}=\frac{a}{v},
\end{equation}
where $a$ is a constant that has dimension of squared speed. Such reciprocal speed $v_{rec}$ will be better understood
later. It is interesting to notice that a similar idea of considering an ``internal motion" (internal degree of freedom) 
for microparticles was thought by Natarajan\cite{32}. 

In addition to (10) and (11), we already know that, at the non-Galilean frame $S_0$, we should have the condition of equality of the time
 intervals, namely $\Delta t=\Delta\tau$ for $v=v_0$. In accordance with (10), this occurs only if

\begin{equation}
\frac{[f(v_0)]^2}{c^2}=\frac{v_0^2}{c^2}\Rightarrow f(v_0)=v_0.
\end{equation}

 Inserting the condition (12) into (11), we find

\begin{equation}
a=v_0^2
\end{equation}

  And so we obtain

\begin{equation}
f(v)=v_{int}= v_{rec}=\frac{v_o^2}{v}
\end{equation}

According to (14) and also considering the cases (i) and (ii), we observe respectively that $f(c)=V=v_0^2/c$ ($V$ is the reciprocal
speed of $c$) and $f(V)=c=v_0^2/V$ ($c$ is the reciprocal speed of $V$), from where we find

\begin{equation}
v_0=\sqrt{cV}
\end{equation}

 Inserting (15) into (14) and after into (10), we finally obtain

\begin{equation} 
\Delta\tau\sqrt{1-\frac{\it{V}^2}{v^2}}=\Delta t\sqrt{1-\frac{v^2}{c^2}},
\end{equation}
being $\alpha=f(v)/c=v_{rec}/c=V/v$ and $\beta=v/c$. In fact, if $v=v_0=\sqrt{cV}$ in (16), then we have $\Delta\tau =\Delta t$. Therefore,
 we conclude that $S_0(v=v_0)$ is the improper non-Galilean reference frame of SSR, so that, if

a) $v>>v_0$ ($v\rightarrow c$)$\Rightarrow\Delta t>>\Delta\tau$: It is the well-known {\it improper time dilatation} of SR.

b) $v<<v_0$ ($v\rightarrow V$) $\Rightarrow\Delta t<<\Delta\tau$: Let us call this new result {\it contraction of improper time}\cite{10}.

 In SSR it is interesting to notice that we recover the newtonian regime of speeds only if $V<<v<<c$, which represents an intermediary
regime where $\Delta\tau\approx\Delta t$.

  Substituting (14) in (9) and also considering (15), we obtain

\begin{equation}
\overline{O^{\prime}C}=\Delta x^{\prime}_v=v_{rec}\Delta\tau=\frac{v_0^2}{v}\Delta\tau=\frac{cV}{v}\Delta\tau=\alpha c\Delta\tau
\end{equation}

  We verify that, if $V\rightarrow 0$ or $v_0\rightarrow 0\Rightarrow\overline {O^{\prime}C} = \Delta x^{\prime}_v=0$,
 restoring the classical case of SR where there is no motion in reciprocal space, i.e., $v_{rec}=0$. And also, if $v>>v_0\Rightarrow\Delta
 x^{\prime}_v\approx 0$, i.e., we get an approximation where $v_{rec}$ can be neglected. We also verify that, if $v=v_0$, this implies
$\Delta x^{\prime}_{[v=v_0]}=v_0\Delta\tau$, as we have already obtained from (8) with the condition $\Delta t=\Delta\tau$ ($v=v_0$).
 This is the unique condition where we find $v=v_{rec}=v_0$.

  From (17), it is also important to notice that, if $v\rightarrow c$, we have $\Delta x^{\prime}(c)=V\Delta\tau$ and, if $v\rightarrow V$
 ($S_V$), we have $\Delta x^{\prime}(V)=c\Delta\tau$. This means that, when the particle momentum increases ($v\rightarrow c$), such a
 particle becomes much better localized upon itself ($V\Delta\tau\rightarrow 0$); and when its momentum decreases ($v\rightarrow V$), it
 becomes completely delocalized because it gets closer to the non-local ultra-referential $S_V$ where $\Delta
 x^{\prime}_v=\Delta x^{\prime}_{max}=c\Delta\tau\rightarrow\infty$. That is the reason why we realize that speed $v$ (momentum) and
 position (delocalization $\Delta x^{\prime}_v=v_{rec}\Delta\tau$) operate mutually like reciprocal quantities in SSR since we have
 $\Delta x^{\prime}_v\propto v_{rec}\propto v^{-1}$ (see (11) or (14)). This provides a basis for the fundamental comprehension
 of the quantum uncertainties emerging from the space-time of SSR. 

 It is interesting to observe that we may write $\Delta x^{\prime}_v$ in the following way:

\begin{equation}
\Delta x^{\prime}_v = \frac{(V\Delta\tau)(c\Delta\tau)}{v\Delta\tau}
\equiv\frac{\Delta x^{\prime}_5\Delta x^{\prime}_4}{\Delta x^{\prime}_1},
\end{equation}
where $V\Delta\tau=\Delta x^{\prime}_5$, $c\Delta\tau=\Delta x^{\prime}_4$ and $v\Delta\tau=\Delta x^{\prime}_1$. We know that
 $c\Delta t=\Delta x_4$ and $v\Delta t=\Delta x_1$. So we rewrite (8), as follows:

\begin{equation}
\Delta x^{\prime 2}_4-\frac{\Delta x^{\prime 2}_5\Delta x^{\prime 2}_4}{\Delta x^{\prime 2}_1}=\Delta x_4^2 - \Delta x_1^2
\end{equation}

 If $\Delta x^{\prime}_5\rightarrow 0$ ($V\rightarrow 0$), this implies $\Delta x^{\prime}_v=0$. So we recover the invariance of the
 4-interval in Minkowski space-time, namely $\Delta S^2=(\Delta x_4^2-\Delta x_1^2)=\Delta S^{\prime 2}=\Delta x^{\prime 2}_4$.

 As we have $\Delta x^{\prime}_v>0$, we observe that $\Delta S^{\prime 2}=\Delta x^{\prime 2}_4>\Delta S^2=(\Delta x_4^2-\Delta x_1^2$).
 Thus we may write (19), as follows:

\begin{equation}
\Delta S^{\prime 2} = \Delta S^2 + \Delta x^{\prime 2}_v,
\end{equation}
where $\Delta S^{\prime}=\overline {AC}$, $\Delta x^{\prime}_v=\overline {O^{\prime}C}$ and $\Delta S=\overline {AO^{\prime}}$
(refer to figure 3).

 For $v>>V$ or $v\rightarrow c$, we have $\Delta S^{\prime}\approx\Delta S$, hence the angle $\theta\approx\frac{\pi}{2}$ (figure 3). 
In the approximation for the macroscopic world (large masses), we get $\Delta x^{\prime}_v=\Delta x^{\prime}_5=0$ (hidden dimension); so
 $\theta=\frac{\pi}{2}\Rightarrow\Delta S^{\prime}=\Delta S$ ($V=0$).

   For $v\rightarrow V$, we would have $\Delta S^{\prime}>>\Delta S$, so that $\Delta S^{\prime}\approx\Delta x^{\prime}_v\approx
 c\Delta\tau\rightarrow\infty$ since $\Delta\tau\rightarrow\infty$ and $\theta\rightarrow\pi$. In this new relativistic limit
 (ultra-referential $S_V$), due to the complete delocalization $\Delta x^{\prime}_v\rightarrow\infty$, the 4-interval $\Delta S^{\prime}$
 loses completely its equivalence with respect to $\Delta S$ because we have $\Delta x^{\prime}_5\rightarrow\infty$ (see (19)).

    Equation (20) (or (19)) shows us a break of $4$-interval invariance ($\Delta S^{\prime}\neq\Delta S$), which becomes noticeable
 only in the limit $v\rightarrow V$ (close to $S_V$, i.e., $\Delta x^{\prime}_v\rightarrow\infty$). However, a new invariance is restored
 when we implement an effective intrinsic dimension ($x^{\prime}_5$) for the moving particle at its non-Galilean frame $S^{\prime}$ by
 means of the definition of a new interval, namely:

\begin{equation}
\Delta S_5=\Delta x^{\prime}_4\sqrt{1-\frac{\Delta x^2_5}{\Delta x^{\prime 2}_1}}=c\Delta\tau\sqrt{1-\frac{V^2}{v^2}},
\end{equation}
so that $\Delta S_5\equiv\Delta S=\sqrt{\Delta S^{\prime 2}-\Delta x^{\prime 2}_v}$ (see (20)).

 We have omitted the index $\prime$ of $\Delta x_5$, as such interval is given only at the non-Galilean proper frame ($S^{\prime}$),
being an intrinsic (proper) dimension of the particle. However, from a practical viewpoint, namely for experiences of higher energies, the
electron approximates more and more to a punctual particle since $\Delta x_5$ becomes hidden.

 Actually the new interval $\Delta S_5$, which could be simply denominated as an effective 4-interval
 $\Delta S=c\Delta\tau*=c\Delta\tau\sqrt{1-\alpha^2}$, guarantees the existence of a certain non-null internal dimension of the particle
 (see (21)), which leads to $\Delta x^{\prime}_v>0$ and thus $v_{rec}\neq 0(>V)$.

Comparing (21) with the left-hand side of equation (16), we may alternatively write

\begin{equation}
\Delta t=\Psi\Delta\tau=\frac{\Delta S_5}{c\sqrt{1-\frac{v^2}{c^2}}}=
\Delta\tau\frac{\sqrt{1-\frac{V^2}{v^2}}}{\sqrt{1-\frac{v^2}{c^2}}} ,
\end{equation}
where $\Delta S_5$ corresponds to the invariant effective 4-interval, i.e., $\Delta S_5\equiv\Delta S$
(segment $\overline{AO^{\prime}}$ in figure 3).

Inserting (17) into (8), we obtain

\begin{equation}
c^2\Delta\tau^2=\frac{1}{\left(1-\frac{V^2}{v^2}\right)}[c^2\Delta t^2-v^2\Delta t^2],
\end{equation}
 where we get $dS_v^2=\theta(v)^{-2}g_{\mu\nu}dx^{\mu}dx^{\nu}$\cite{33}\cite{34}\cite{35}\cite{36}. 

By placing eq.(23) in a differential form and manipulating it, we will obtain

\begin{equation}
c^2\left(1-\frac{V^2}{v^2}\right)\frac{(d\tau)^2}{(dt)^2} + v^2 = c^2
\end{equation}

We may write (24) in the following alternative way:

\begin{equation}
\frac{(dS_5)^2}{(dt)^2} + v^2 = c^2,
\end{equation}
where $dS_5=\theta(v)cd\tau=c\sqrt{1-\frac{V^2}{v^2}}d\tau$. 

 Equation (24) shows us that the speed related to the marching of the time (``temporal-speed''), that is
 $v_t=c\sqrt{1-\frac{V^2}{v^2}}\frac{d\tau}{dt}$, and the spatial speed $v$ with respect to the background field ($S_V$) form the
legs of a rectangle triangle according to figure 4.

\begin{figure}
\includegraphics[scale=0.8]{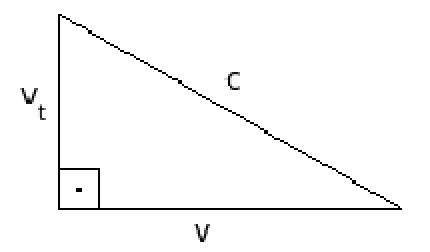}
\caption{We have $c=(v_t^2+v^2)^{1/2}$, which represents the space-temporal speed of any particle (hypothenuse of the triangle=$c$).
 The novelty here is that such a structure of space-time implements the ultra-referential $S_V$. This implementation arises at the vertical
 leg $v_t$.}
\end{figure}

  In accordance with figure 4, we should consider three important cases, namely:

 a)    If $v\approx c$, then $v_t\approx 0$, where $\Psi>>1$, since $\Delta t>>\Delta\tau$ (dilatation of improper time).

 b)    If $v=v_0$, then $v_t=\sqrt{c^2-v_0^2}$, where $\Psi=\Psi_0=\Psi(v_0)=1$, since $\Delta t=\Delta\tau$.

 c)    If $v\approx V$, then $v_t\approx\sqrt{c^2-V^2} =c\sqrt{1-\xi^2}$, where $\Psi<<1$, since $\Delta t<<\Delta\tau$ (dilatation of
 the proper time (at $S^{\prime}$) with respect to the improper one (at $S_0$)).

\begin{figure}
\begin{center}
\includegraphics[scale=0.6]{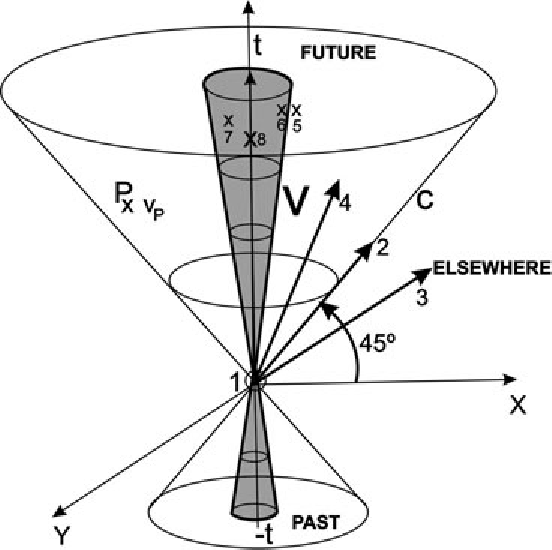}
\end{center}
\caption{The external and internal conical surfaces represent respectively the speed $c$ and $V$, where $V$ is represented by the
 dashed line, namely a definitely prohibited boundary. For a point $P$ in the interior of the two conical surfaces, there is a corresponding
 internal conical surface so that $V<v_p\leq c$, where $v_p$, represented by the horizontal leg of the triangle in figure 4, is associated
 with a temporal speed $v_{tp}=\sqrt{c^2-v_p^2}$ (the vertical leg of the triangle) given in the representation for {\it temporal cone} 
(figure 6). The $4$-interval $S_4$ is of type time-like. The $4$-interval $S_2$ is a light-like interval (surface of the light cone). 
The $4$-interval $S_3$ is of type space-like (elsewhere). The novelty in spacetime of SSR are the $4$-intervals $S_5$ (surface of
the dark cone) representing an infinitly dilated time-like interval\cite{37}, including the $4$-intervals $S_6$, $S_7$ and 
$S_8$ inside the dark cone for representing a new region of type space-like\cite{37}.}
\end{figure}

\begin{figure}
\begin{center}
\includegraphics[scale=0.6]{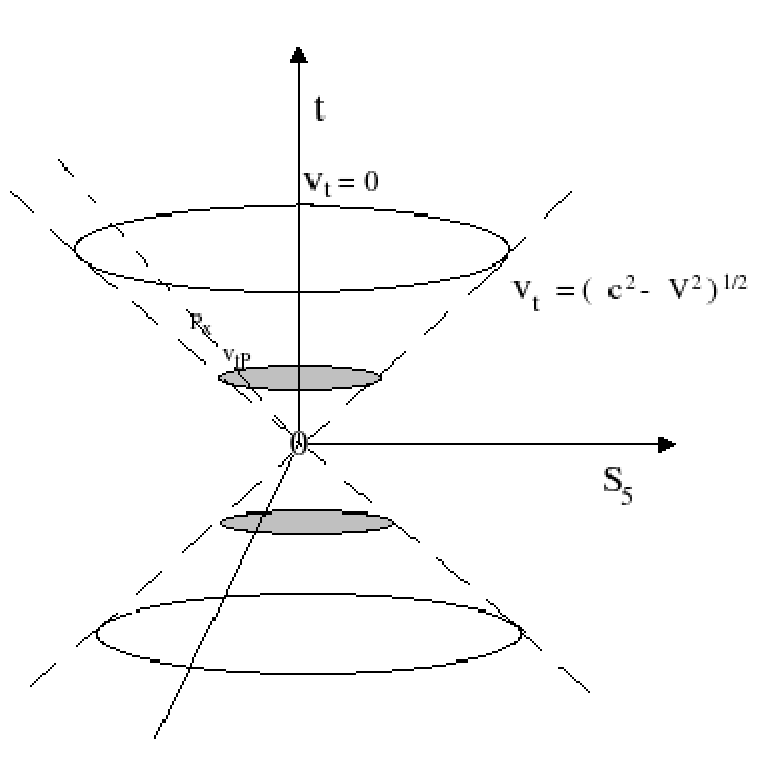}
\end{center}
\caption{Comparing this figure with figure 5, we notice that the dashed line on the internal cone of figure 5 ($v=V$) corresponds to
 the dashed line on the surface of the external cone of this figure, where $v_t=\sqrt{c^2-V^2}$, which represents a definitely forbidden
 boundary in this cone representation of temporal speed $v_t$. On the other hand, $v=c$ (photon) is represented by the solid line of 
figure 5, which corresponds to the temporal speed $v_t=0$ in this figure. In short, we always have $v^2+v_t^2=c^2$, being $v$ given
in the spatial (light) cone (figure 5) and $v_t$ for the temporal cone represented in this figure.}
\end{figure}

 Based on equation (23) or also by inserting (17) into (8), we obtain

  \begin{equation}
  c^2\Delta t^2 - v^2\Delta t^2 = c^2\Delta\tau^2 -
 \frac{v_0^4}{v^2}\Delta\tau^2
  \end{equation}

 In eq.(26), when we transpose the 2nd term from the left-hand side to the right-hand side and divide the equation by $(\Delta t)^2$, we
 obtain (24) given in differential form. Now, it is important to observe that, upon transposing the 2nd term from the right-hand side to
 the left-hand one and dividing the equation by $(\Delta\tau)^2$, we obtain the following equation in differential form, namely:

  \begin{equation}
  c^2\left(1-\frac{v^2}{c^2}\right)\frac{(dt)^2}{(d\tau)^2} + \frac{v_0^4}{v^2}=c^2
  \end{equation}

 From (21) and (16), we obtain $dS_5=cd\tau\sqrt{1-\alpha^2}=cdt\sqrt{1-\beta^2}$. Hence we can write (27) in the following alternative
 way:

  \begin{equation}
  \frac{(dS_5)^2}{(d\tau)^2} + \frac{v_0^4}{v^2}= c^2
  \end{equation}

  Equation (27) (or (28)) reveals a complementary way of viewing equation (24) (or (25)), which takes us to that idea of
 reciprocal space for conjugate quantities. Thus let us write (27) (or (28)) in the following way:

  \begin{equation}
  v_{trec}^2 + v_{rec}^2 =c^2,
  \end{equation}
where $v_{trec}(=(v_t)_{int}=c\sqrt{1-\frac{v^2}{c^2}}\frac{dt}{d\tau})$ represents an internal (reciprocal)``temporal speed"
 and $v_{rec}(=v_{int}=f(v)=\frac{v_0^2}{v})$ is the internal (reciprocal) spatial speed. Therefore we can represent a rectangle
 triangle which is similar to that of figure 4, but now being represented in a reciprocal space. For example, if we assume $v\rightarrow c$
 (equation (24)),
we obtain $v_{rec}=lim._{v\rightarrow c}f(v)\rightarrow V$ (equation (27)). For this same case, we have $v_t\rightarrow 0$ (equation (24))
 and $v_{trec}=\frac{dS_5}{d\tau}\rightarrow\sqrt{c^2-V^2}$ (equation (27) or (28)). On the other hand, if $v\rightarrow V$ (eq.(24)), we
 have $v_{rec}\rightarrow\frac{v_0^2}{V}=c$ (eq.(27)), where $v_t\rightarrow\sqrt{c^2-V^2}$ (eq.(24)) and $(v_t)_{int}=v_{trec}\rightarrow
 0$ (eq.(27)). Thus we should observe that there are altogheter four cone representations in SSR, namely:

\begin{equation}
 spatial~representations:\left\{\begin{array}{ll}
a_1)  v=\frac{dx}{dt}, in~eq.(24),\\
represented~in~Fig.5;\\
 b_1) v_{rec}=\frac{d x^{\prime}_v}{d\tau}=\frac{v_0^2}{v},\\
 in~eq.(27).
\end{array}
 \right.
\end{equation}

\begin{equation}
 temporal~representations:\left\{\begin{array}{ll}
a_2) v_t=c\sqrt{1-\frac{V^2}{v^2}}\frac{d\tau}{dt}\\
=c\sqrt{1-\frac{v^2}{c^2}}, in~eq.(24),\\
represented~in~Fig.6;\\

b_2) v_{trec}=c\sqrt{1-\frac{v^2}{c^2}}\frac{dt}{d\tau}\\
=c\sqrt{1-\frac{V^2}{v^2}}, in~eq.(27).
\end{array}
 \right.
 \end{equation}

The chart in figure 7 shows us the four cone representations in the space-time of SSR.

\begin{figure}
\begin{center}
\includegraphics[scale=0.55]{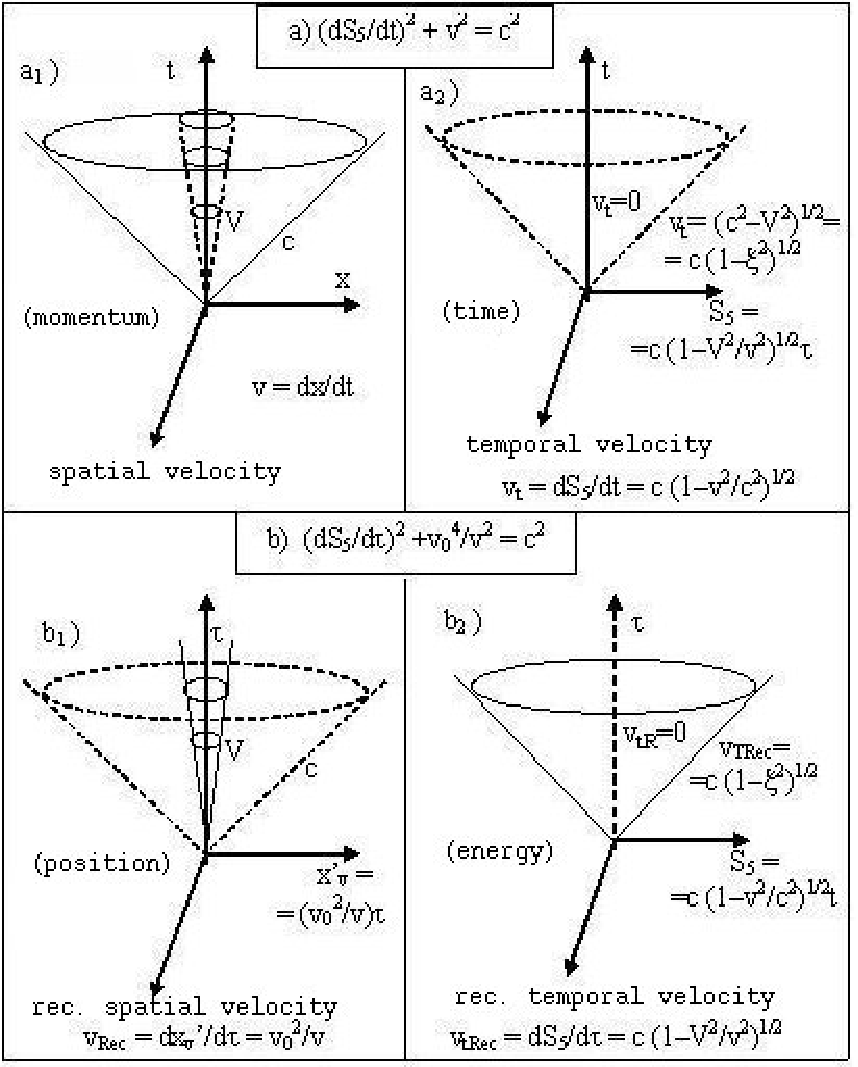}
\end{center}
\caption{The spatial representations in $a_1$ (also shown in figure 5) and $b_1$ are related respectively to velocity $v$ (momentum) and
 position (delocalization $\Delta x^{\prime}_v=f(v)\Delta\tau=v_{int}\Delta\tau=v_{rec}\Delta\tau=(v_0^2/v)\Delta\tau$), which
 represent conjugate (reciprocal) quantities in the space. On the other hand, the temporal representations in $a_2$ (also shown in figure
 6) and $b_2$ are related respectively to the time ($\propto v_t$) and the energy ($\propto v_{trec}=(v_t)_{int}\propto v_t^{-1}$), which
 represent reciprocal quantities in the time. Hence we can perceive that such four cone representations of SSR provide a basis
 for the fundamental understanding of the uncertainty relations.}
\end{figure}

Now, by considering (16),(22) and (31), looking at $a_2$ and $b_2$ in figure 7, we may observe

 \begin{equation}
 \Psi^{-1}=\frac{\Delta\tau}{\Delta t}= \frac{\sqrt{1-\frac{v^2}{c^2}}}{\sqrt{1-\frac{V^2}{v^2}}}=
\frac{v_t}{c\sqrt{1-\frac{V^2}{v^2}}}= \frac{v_t}{v_{trec}}\propto (time)
  \end{equation}

  and

  \begin{equation}
\Psi=\frac{\Delta t}{\Delta\tau}= \frac{\sqrt{1-\frac{V^2}{v^2}}}{\sqrt{1-\frac{v^2}{c^2}}}=
\frac{v_{trec}}{c\sqrt{1-\frac{v^2}{c^2}}}=\frac{v_{trec}}{v_t}\propto E,
  \end{equation}
being $E=Energy\propto (time)^{-1}$.

 From (33), as we have $E\propto\Psi$, we obtain $E=E_0\Psi$, where $E_0$ is a constant. Hence, by considering
 $E_0=m_0c^2$, we write

\begin{equation}
E= m_0c^2\frac{\sqrt{1-\frac{V^2}{v^2}}}{\sqrt{1-\frac{v^2}{c^2}}},
\end{equation}
where $E$ is the total energy of the particle with respect to the ultra-referential $S_V$ of the background field. In (33) and (34), we
 observe that, if $v\rightarrow c\Rightarrow E\rightarrow\infty$ and $\Delta\tau\rightarrow 0$ for $\Delta t$ fixed; if $v\rightarrow
 V\Rightarrow E\rightarrow 0$ and $\Delta\tau\rightarrow\infty$, also for $\Delta t$ fixed. If $v=v_0=\sqrt{cV}\Rightarrow E=E_0=m_0c^2$
 (energy of ``quantum rest'' at $S_0$).

  Figure 8 shows us the graph for the energy $E$ in eq.(34).

\begin{figure}
\begin{center}
\includegraphics[scale=0.7]{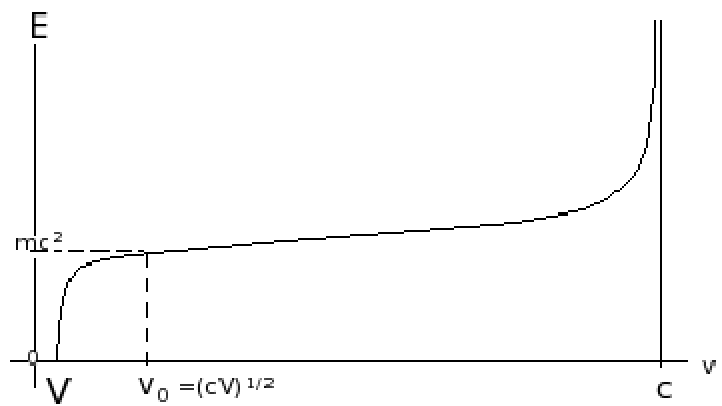}
\end{center}
\caption{$v_0$ represents the speed in relation to $S_V$, from where we get the proper energy of the particle ($E_0=m_0c^2$), being
 $\Psi_0=\Psi(v_0)=1$. For $v<<v_0$ or closer to $S_V$ ($v\rightarrow V$), a new relativistic correction on energy arises, so that
 $E\rightarrow 0$.}
\end{figure}

 The momentum of the particle with respect to $S_V$ is given as follows:

\begin{equation}
 P = m_0 v\frac{\sqrt{1-\frac{V^2}{v^2}}}{\sqrt{1-\frac{v^2}{c^2}}},
\end{equation}
For $v=v_0$, we find $P=m_0v_0=m_0\sqrt{cV}$, as $\Psi(v_0)=1$. If $v\rightarrow V\Rightarrow P\rightarrow 0$.

Some other aspects of such a relativistic dynamics in SSR were investigated in more details in a previous paper\cite{10}.

 As the momentum $P$ of a particle (eq.35) is given with respect to the ultra-referential $S_V$, we can also find its de-Broglie
wavelength with respect to $S_V$, namely:

\begin{equation}
\lambda=\frac{h}{P}=\frac{h}{m_0v}\frac{\sqrt{1-\frac{v^2}{c^2}}}{\sqrt{1-\frac{V^2}{v^2}}}
\end{equation}

 According to eq.(36), if we have $v\rightarrow c$, this leads to $\lambda\rightarrow 0$ (spatial contraction), and if
$v\rightarrow V$ (close to the ultra-referential $S_V$), this implies $\lambda\rightarrow\infty$ (spatial dilatation leading to a
complete delocalization).

 \section{\label{sec:level1} The origin of the uncertainty principle}

  The actual momentum of a massive particle with respect to $S_V$ is $P=\Psi m_0 v$, whose conjugate value is 
 $\Delta x^{\prime}_v=v_{rec}\Delta\tau=\frac{v_0^2}{v}\Delta\tau=\frac{v_0^2}{v}\Delta t\Psi^{-1}$, where $\Delta\tau=\Psi^{-1}\Delta t$ 
 (refer to eq.16). Since $\Delta x^{\prime}_v$ represents a delocalization working like an intrinsic uncertainty on position of the
 particle, the momentum $P$, which represents its conjugate value, should be also interpreted as an uncertainty on
 momentum, namely $P=\Delta p$ as we will see soon.

  As $P$ is the actual momentum given with respect to the ultra-referential $S_V$, it is always inaccessible
 to a classical observer at any Galilean reference frame $S$. Due to this impossibility to know exactly the actual momentum $P$
 from any Galilean frame $S$, so $P$ appears as an uncertainty $\Delta p$ on the wave-packet of the particle at the frame $S$. In other
 words this means that the speed $v$ (non-Galilean frame $S^{\prime}$) given with respect to $S_V$ appears as $\Delta v$ at any inertial
 (Galilean) reference frame $S$, i.e., $v=\Delta v$. For example, if $v\rightarrow V$ ($P\rightarrow 0$), this means
 $\Delta v\rightarrow V$ ($\Delta p\rightarrow 0$), and so we have $\Delta x^{\prime}_v\rightarrow\infty$, i.e., the particle is
 completely delocalized in the space.

   Now we can compute the following quantity, namely:

 \begin{equation}
  \Delta x^{\prime}_vP=\frac{v_0^2}{v}\Delta t\Psi^{-1}\Psi m_0 v=(m_0v_0)(v_0\Delta t),
\end{equation}

  In obtaining (37), we have also considered the relations
 $\Delta x^{\prime}_v=\frac{v_0^2}{v}\Delta\tau$, $\Delta\tau=\Delta t\Psi^{-1}$ and $P=\Psi m_0 v$. We have
 $v_0=\sqrt{cV}$. Of course we get $\Delta x^{\prime}_vP=0$ in SR since $V=0$ ($v_0=0$) in the classical space-time.

  As a non-inertial frame has the same state of motion (e.g: acceleration) given for any inertial (Galilean) frame, we also expect that
 a non-Galilean frame of SSR, having an absolute speed $v$ ($V<v<c$) with respect to the preferred frame $S_V$, remains invariant for any
 inertial (Galilean) frame. In other words, we say that all inertial (Galilean) frames must agree among themselves that a given velocity
 $v$ of a non-Galilean referential with respect to the ultra-referential $S_V$ should remain the same for them. However, since such an 
 absolute speed $v$ is inaccessible for us at any Galilean reference frame in the classical space-time, it should appear 
 as a width $\Delta v$ on the wave-packet of the moving particle, i.e., we get $v=\Delta v$ at any inertial (Galilean) reference frame.

  Since the particle has an actual wavelength $\lambda$ given with respect to the background frame $S_V$ (eq.36), it is natural to think
that the delocalization $\Delta x^{\prime}_v$ becomes of the order of its wavelength for a special condition such that $\lambda$
corresponds to a certain intrinsic uncertainty on position $(\Delta x^{\prime}_v)_0$, namely:

\begin{equation}
 \lambda\sim(\Delta x^{\prime}_v)_0=\frac{v_0^2}{v}(\Delta\tau)_0=\frac{v_0^2}{v}(\Delta t)_0\Psi^{-1},
\end{equation}
where $(\Delta t)_0$ will be computed. We have $v_0^2/v=v_{rec}$.

 Now comparing (38) with (36), we write

\begin{equation}
 \lambda=\frac{h}{m_0v}\Psi^{-1}\sim\frac{v_0^2}{v}(\Delta t)_0\Psi^{-1},
\end{equation}
from where we obtain

\begin{equation}
 m_0v_0^2(\Delta t)_0\sim h
\end{equation}

Finally by comparing (40) with (37), we find

\begin{equation}
 (\Delta x^{\prime}_v)_0P=m_0v_0^2(\Delta t)_0\sim h
\end{equation}

or alternatively we write

\begin{equation}
 (\Delta x^{\prime}_v)_0\Delta p=m_0v_0\lambda_0\sim h,
\end{equation}
where we have $P=\Delta p$ at any Galilean reference frame, and from (41) we find $(\Delta t)_0\sim h/m_0v_0^2$
or $v_0(\Delta t)_0\sim\lambda_0=h/m_0v_0$, being $v_0=\sqrt{cV}$. So we write $(\Delta t)_0=\lambda_0/v_0$.

 The relation (42) is the uncertainty relation for momentum emerging from the quantum space-time of SSR.

 Now in order to obtain the uncertainty relation for energy, we first rewrite the relation (40), as follows:

 \begin{equation}
 m_0cV(\Delta t)_0\sim h,
 \end{equation}
where $v_0^2=cV$.

 Alternatively we can write (43) in the following way:

\begin{equation}
 m_0c^2\xi(\Delta t)_0\sim h,
 \end{equation}
where $\xi=V/c$.

From (44), we simply verify that $\xi(\Delta t)_0=Vh/cm_0v_0^2=h/m_0c^2=(\Delta t)_c$, where we have $(\Delta t)_c=\lambda_c/c$, being
$\lambda_c(=h/m_0c)$ the Compton wavelength. So we write $\lambda_c/c=\xi\lambda_0/v_0\Rightarrow\lambda_c=(V/v_0)\lambda_0
=(v_0/c)\lambda_0$.

The relation (44) can be also written as follows:

\begin{equation}
 m_0c^2\Psi(\Delta t)_c\Psi^{-1}\sim h,
 \end{equation}
from where we take $m_0c^2\Psi=E$ (eq.34) and also $(\Delta t)_c\Psi^{-1}=(\Delta\tau)_c$ (see eq.(16) or (22)), being
$\Psi=\sqrt{1-\alpha^2}/\sqrt{1-\beta^2}$. Therefore, from (45) we find

\begin{equation}
 E(\Delta\tau)_c=m_0c^2(\Delta t)_c\sim h,
 \end{equation}

or alternatively we write

\begin{equation}
 \Delta E(\Delta\tau)_c=m_0c\lambda_c\sim h,
 \end{equation}
where $\lambda_c=c(\Delta t)_c$.

 We have considered $P=\Delta p$ at any Galilean (inertial) reference frame. So we should also consider that the energy $E$ with respect
to background frame $S_V$ (eq.34) appears as an uncertainty $\Delta E$ at any Galilean reference frame, i.e., we have $E=\Delta E$.

 According to (47), in the limit $v\rightarrow V$, this implies $\Delta E\rightarrow 0$, leading to 
$(\Delta\tau)_c=(\Delta t)_c\Psi^{-1}_{(v=V)}\rightarrow\infty$.

 The relation (47) is the uncertainty relation for energy emerging from the quantum space-time of SSR.

{\noindent\bf  Acknowledgedments}

 I am grateful to Prof. Jonas Durval Cremasco, Em\'ilio C. Gerra, A. C. Amaro de Faria Jr. and Alisson Xavier for 
interesting discussions.

\end{document}